\documentclass[aps,pra,twocolumn,superscriptaddress,showpacs,aps]{revtex4}
\usepackage{amsmath,epsfig,amsfonts,graphicx}
\usepackage{subfigure}
\usepackage{graphicx}
\usepackage[american]{babel}
\usepackage{amssymb}
\usepackage{amsfonts}
\usepackage{color}
\usepackage[normalem]{ulem}

\begin{document}
\newcommand{\be}{\begin{equation}}
\newcommand{\ee}{\end{equation}}
\newtheorem{corollary}{Corollary}[section]
\newtheorem{remark}{Remark}[section]
\newtheorem{definition}{Definition}[section]
\newtheorem{theorem}{Theorem}[section]
\newtheorem{proposition}{Proposition}[section]
\newtheorem{lemma}{Lemma}[section]
\newtheorem{help1}{Example}[section]
\title{
The dynamical playground of a higher-order cubic Ginzburg-Landau
equation:
from orbital connections and limit cycles to invariant tori and the onset of chaos
}
\author{V. Achilleos}
\affiliation{Laboratoire d' Acoustique de l' Universit\'{e} du Maine, Avenue O. Messiaen 72085, Le Mans, France}
\author{A. R. Bishop}
\affiliation{Center for Nonlinear Studies and Theoretical Division,
Los Alamos National Laboratory, Los Alamos, New Mexico 87545, USA}
\author{S. Diamantidis}
\affiliation{Department of Mathematics, University of the Aegean, Karlovassi, 83200 Samos, Greece}
\author{D. J. Frantzeskakis}
\affiliation{Department of Physics, University of Athens, Panepistimiopolis, Zografos, Athens 15784, Greece}
\author{T. P. Horikis}
\affiliation{Department of Mathematics, University of Ioannina, Ioannina 45110, Greece}
\author{N. I. Karachalios}
\affiliation{Department of Mathematics, University of the Aegean, Karlovassi, 83200 Samos, Greece}
\author{P. G. Kevrekidis}
\affiliation{Department of Mathematics and Statistics, University of Massachusetts, Amherst MA 01003-4515, USA}
\affiliation{Center for Nonlinear Studies and Theoretical Division,
Los Alamos National Laboratory, Los Alamos, New Mexico 87545, USA}

\begin{abstract}
The dynamical behavior of a higher-order cubic Ginzburg-Landau equation is found to include
a wide range of scenarios due to the interplay of higher-order physically relevant terms. 
We find that the competition between the third-order dispersion and
stimulated Raman scattering effects, gives rise to rich dynamics: this
extends from
Poincar\'{e}-Bendixson--type scenarios, in the sense that bounded solutions may converge either to distinct equilibria via orbital connections, or space-time periodic solutions, to the emergence of almost periodic and chaotic behavior. 
One of our main results is that the third-order dispersion has a dominant role
in the development of such complex dynamics, since it can be
chiefly responsible
(i.e., even in the absence of the other higher-order effects) for the existence
of the  periodic, quasi-periodic and chaotic spatiotemporal structures.
Suitable low-dimensional phase space diagnostics are devised and used to illustrate the different possibilities and identify
their respective parametric intervals over multiple parameters of the
model.
\end{abstract}


\maketitle


\section{Introduction}
Nonlinear evolution equations are often
associated with the theory of solitons and integrable systems \cite{Ablo}. A prime example is
the nonlinear Schr\"odinger (NLS) equation which constitutes one of the universal
nonlinear evolution equations, 
with applications ranging from deep water waves to optics \cite{ablo2}. 
Remarkable phenomena are also exhibited by its higher-order variants, 
emerging in a diverse spectrum of applications, such as nonlinear
optics \cite{KodHas87}, nonlinear metamaterials \cite{p31}, and water waves in finite depth \cite{johnson,sedletsky,slun}. On the other hand, dissipative variants
of NLS models incorporating gain and loss have also been used in optics
\cite{akbook}, e.g., in the physics of mode-locked lasers \cite{laser1,laser2}
(see also the relevant works \cite{tsoy,tsoy2})
and polariton superfluids
\cite{daniele} -- see, e.g., Ref.~\cite{akbook2} for various applications.
Note that such dissipative NLS models can be viewed as variants
of the complex Ginzburg-Landau
(CGL) equation, which has been extensively studied, especially in
the context of pattern formation in far-from-equilibrium systems \cite{ar}.

Dissipative nonlinear evolution equations (incorporating gain, loss, external driving,
or combinations thereof) may exhibit
(and potentially be attracted to) low-dimensional
dynamical features, such as: (a) one or more equilibria (and
orbits connecting them), (b) periodic orbits, (c)
quasi-periodic orbits or (d) low-dimensional
chaotic dynamics~\cite{smale}.
%
The availability of the 
dynamical scenarios (a)-(d) depends on
the effective dimensionality of the low dimensional behavior; one-dimensionality
only allows fixed points, planar systems governed by the
Poincar{\'e}-Bendixson (PB) theorem~\cite{smale}
can also feature periodic orbits, while higher dimensions allow for 
quasi-periodic or chaotic dynamics.
Various prototypical partial differential equation models
have demonstrated
a PB-type behavior as an intermediate bifurcation stage in the route to spatiotemporal chaos. 
Examples include the Kuramoto-Sivashinsky \cite{Cv1}
and complex Ginzburg-Landau (CGL) equations; 
regarding the CGL model, which is of primary interest in this work,
we refer to the seminal works \cite{NBGL} for the spatiotemporal transition to chaos. 
In addition to the above autonomous systems, spatiotemporal chaos was also found in non-autonomous ones, due to the interplay between loss and external forces, such as
the damped-driven NLS
\cite{nobe,kai,eli} (where the hyperbolic structure of the underlying integrable NLS is a prerequisite \cite{Li}) 
and the sine-Gordon \cite{Sin1} system.

In this work, 
we focus on the role of higher order effects, 
and investigate the possibility of bifurcation phenomena leading to the existence of  
the above prototypical examples of low-dimensional dynamics 
in an autonomous, physically important higher-order CGL-type model.
This model, is motivated by the higher-order NLS equation that is commonly
used, e.g., in studies of ultrashort pulses in optical fibers \cite{KodHas87},
but also incorporates (linear or nonlinear) gain and loss; it is, thus, a
physically relevant variant of a higher-order cubic CGL equation -- without the diffusion term.
Note that higher-order versions of the CGL equation, have only recently started attracting attention \cite{LatasA}, while extended second-order CGL models have been extensively studied
in various contexts previously~\cite{akbook,ar,akbook2}. 
In particular, we refer to the pioneering work \cite{laser2}, followed by the
important contributions \cite{tsoy,tsoy2}, which revealed the existence of the
aforementioned low-dimensional dynamical scenarios for second-order quintic CGL models.
The results of \cite{laser2,tsoy,tsoy2} were established by numerical and even
analytical reductions
to suitable finite dimensional dynamical systems, capturing the long-time dynamics of the
original infinite dimensional one. Notably, the revealed dynamical scenarios were
associated with a variety of novel localized structures (known as pulsating solitons). However, a crucial feature of these models as acknowledged, e.g.,
by the authors of~\cite{laser2} was the presence of a higher order
(quintic) nonlinearity. This is a key trait distinguishing that set
of works from the present one where only cubic nonlinearity is employed,
yet the presence of higher order effects, most notably
third order dispersion as we will see below, plays a catalytic role in the
emergence of the relevant phenomenology.


More specifically, we should point out that the {\it third-order cubic} CGL model that we consider herein
is essentially different from the {\it second-order cubic-quintic} model discussed in \cite{laser2,tsoy,tsoy2}, not only from a mathematical but also from a physical point of view:
indeed, in the context of optics, the model considered in the latter works refers to propagation
of short pulses, in the picosecond regime, in media featuring saturation of the nonlinear refractive
index, while the model we consider here is relevant to propagation of
{\it ultrashort} pulses in the sub-picosecond or femtosecond regimes \cite{KodHas87}.
For this reason, our model includes third-order dispersion
and higher-order nonlinear effects, that appear naturally as higher-order corrections of the usual NLS model
in the framework of the reductive perturbation method. In that regard, and if gain and loss are also
incorporated,
%
%
it is important to ask if, and how, the physically important (in the femtosecond time-scale)
higher-order effects may be responsible for tracing a path to complex dynamics,
as a result of the potential breaking of the homoclinic structure of the unperturbed NLS counterpart.

The main findings of our investigations are the following. First, we show that the
incorporation of the gain and loss terms gives rise to the
existence of an attractor; a rigorous proof is provided, based on the interpretation
of the energy balance equation and properties of the functional (phase) space
on which the problem defines an infinite-dimensional flow. The structure of the attractor
is then investigated numerically. Given that our model is characterized by six free
parameters (which renders a systematic investigation of their role a nontrivial task),
we opt to keep four parameters fixed, with values suggested by the physics
of ultrashort optical pulses \cite{KodHas87}, and vary the remaining two. In particular,
we vary the coefficients of the third-order dispersion and the higher-order nonlinear
dissipation, accounting for the stimulated Raman scattering (SRS) effect
(more important reasons for this choice will become apparent below).
We find that, for
sufficiently small SRS coefficient, variations of the third-order dispersion strength
give rise to a transition path 
from dynamics reminiscent of PB, including orbital connections between steady states of high multiplicity and 
convergence to limit cycles, to invariant tori or even chaotic attractors. However,
when the SRS effect becomes stronger, the above scenarios are screened by
convergence to steady-states. 
It is highlighted that the third-order dispersion is found to be
chiefly  responsible
for a dynamical transition from periodic, to quasi-periodic and eventually to chaotic structures.
Therefore, our results show that higher (third)-order dispersion and dissipative (SRS) effects
are important mechanisms for the emergence of complex spatiotemporal transitions in CGL models.
%

Our presentation is organized as follows. In Section~II, we present the model,
and discuss the existence of a  limit set (attractor). Details on the proof of such a limit set are given in the Appendix \ref{apLS}. The structure of the attractor is then
investigated numerically in Section ~III. We thus reveal the emergence of all
dynamical scenarios and corresponding regimes of complex asymptotic behavior.
Finally, Section~IV summarizes our findings.


\section{Motivation and presentation of the model}

Our model 
is motivated by the following higher-order NLS equation:
 \begin{align}
 \label{eq0}
 {\rm i}\partial_t u - \frac{s}{2} \partial_x^2 u + |u|^2u
 &={\rm i}\beta \partial_x^3 u+{\rm i}\mu \partial_x (|u|^2u)
  \nonumber\\
  &+ \left(\sigma_R+{\rm i}\nu\right)u \partial_x(|u|^2),
%
 \end{align}
where $u(x,t)$ is a complex field,
subscripts denote partial differentiation,
$\beta$, $\mu$, $\nu$ and $\sigma_R$ are positive constants, while $s = \pm 1$
denotes normal (anomalous) group velocity dispersion.
Note that Eq.~(\ref{eq0}) can be viewed as a perturbed NLS equation, with the
perturbation (in case of small values of relevant coefficients) appearing in the right-hand side
(see, e.g., Refs.~\cite{KodHas87} and discussion below).

Variants of Eq.~(\ref{eq0}) appear in a variety of physical contexts,
where they are derived at higher-order approximations of perturbation theory [the
lowest-order nonlinear model is simply the NLS equation in the left-hand side
of Eq.~(\ref{eq0})].
The most prominent example is probably 
that of nonlinear optics \cite{KodHas87}. In this case, $t$ and $x$ denote propagation distance
and retarded time,
respectively, while $u(x,t)$
is the 
electric field envelope.
%
%
While the unperturbed NLS equation is sufficient to describe optical pulse propagation,
for ultra-short pulses 
third-order dispersion and self-steepening (characterized by coefficients $\beta$, $\mu$ and $\nu$,
respectively) become important and have to be incorporated in the model.
Similar situations also occur in other contexts and, thus,
corresponding versions of Eq.~(\ref{eq1}) have been derived and used,
e.g., in nonlinear metamaterials \cite{p31}, but also in water waves in finite depth
\cite{johnson,sedletsky,slun}.
Moreover, in the context of optics, and for relatively long propagation distances,
higher-order nonlinear dissipative effects, such
as the 
SRS effect, of strength $\sigma_R>0$, are also important  \cite{KodHas87}.

In addition to the above mentioned effects, our aim is to investigate the dynamics
in the framework of Eq.~(\ref{eq0}), but also incorporating linear or nonlinear gain and loss.
This way, in what follows, we are going to analyze the following model:
 \begin{eqnarray}
 \label{eq1}
\!\!\!\!\!\!\!\!\!
 {\rm i}\partial_t u - \frac{s}{2} \partial_x^2 u + |u|^2u
 &=&{\rm i}\gamma u+{\rm i}\delta|u|^2u
  + {\rm i}\mu \partial_x (|u|^2u)
  \nonumber\\
  &+&{\rm i}\beta \partial_x^3 u+\left(\sigma_R+{\rm i}\nu \right)u \partial_x(|u|^2),
%
 \end{eqnarray}
which includes
linear loss ($\gamma<0$) [or gain ($\gamma>0$)]. These effects are physically
relevant in 
nonlinear optics \cite{KodHas87,akbook,akbook2}: indeed,
nonlinear gain ($\delta>0$) [or loss ($\delta<0$)] may be used to counterbalance the
effects from the linear loss/gain mechanisms and can potentially
stabilize optical solitons -- see, e.g., Refs.~\cite{p35,DJFTheo}.
As is also explained below, here we focus on the case of linear gain, $\gamma>0$, and
nonlinear loss, $\delta<0$, corresponding to a constant gain distribution, and
the intensity-dependent two-photon absorption, respectively (see, e.g., Refs.~\cite{Chen}).

Obviously, the presence of gain/loss renders Eq.~(\ref{eq1}) a higher-order cubic CGL equation (cf. recent studies \cite{LatasA} on such models). 
Note that in Eq.~(\ref{eq1}), diffusion is absent:
such a linear term would be of the form
$\mathrm{i} D\partial_x^2 u$ ($D=$const.), and would appear in the right-hand side of Eq.~(\ref{eq1})
to account for the presence of spectral filtering or linear parabolic gain ($D>0$)
or loss ($D<0$) \cite{laser2,tsoy,tsoy2}. Instead, the equation only
features linear dispersion through the term proportional to $s$ in the
left hand side of Eq.~(\ref{eq1}).

The gain/loss effects are pivotal for the dissipative nature of the infinite-dimensional flow
that will be defined below. This dissipative nature is reflected in the existence of an attractor,
capturing its long-time dynamics; nevertheless, as we will show below, the structure
of the attractor is determined by the remaining higher-order effects.

Here, we focus on the case $s=1$, and supplement Eq.~(\ref{eq1}) with periodic boundary
conditions for $u$ and its spatial derivatives up to the-second order, namely:
\begin{equation}\label{eq2}	
\begin{array}{cc}
 u(x+2L,t)=u(x,t),\;\;\mbox{and}\\
\;\;\;\;\;\;\;\partial^j_x(x+2L,t)=\partial^j_x(x,t),\;\; j=1,2,
\end{array}
\end{equation}
$\forall\;\;(x,t)\in{\mathbb R}\times [0,T]$, for some $T>0$, where $L>0$ is given. The initial condition
\begin{equation}\label{eq3}
u (x,0)=u_0(x),\quad \forall\,x\in{\mathbb R},
\end{equation}
also satisfies the periodicity conditions~(\ref{eq2}). Here, we should mention that the periodic boundary
conditions that we consider here are also motivated by the context of optics. Recalling that the roles
of space and time are interchanged in the latter context, we note that, indeed, in
optical cavities (e.g., ones for lasers), the period $L$ would account for the (retarded)
time it takes light to traverse to the laser cavity once and, thus,
the boundaries represent the same point in the real space-time
(see, e.g., Refs.~\cite{Salin,Haus2,VanWig,AbDT}).
In this context, the dynamics that will be analyzed below
are relevant to the dynamical transitions and the observation of chaotic optical waveforms in fiber ring lasers \cite{VanWig}.

As shown in Ref.~\cite{PartI}, all possible regimes  except $\gamma>0$, $\delta<0$, are associated with finite-time collapse or decay. Furthermore, a critical value $\gamma^*$ can be identified in the regime $\gamma<0$, $\delta>0$, which separates finite-time collapse from the decay of solutions. On the other hand, for $\gamma>0$, $\delta<0$, we prove in Appendix \ref{apLS} the existence of a limit set (attractor) $\omega(u_0)$, attracting all bounded orbits initiating from arbitrary, appropriately smooth initial data $u_0$ (considered as elements of a suitable Sobolev space). In the next Section, 
we will show numerically that the attractor $\omega(u_0)$ captures the full route from PB-type dynamics 
to quasi-periodic or chaotic dynamics.
\section{Numerical results}

The structure of the limit set $\omega(u_0)$,  is
investigated by numerical integration via a high-accuracy pseudo-spectral method.
In our simulations, we fix the half length of $\Omega$ to $L=50$, and
the ratio $-\gamma/\delta$ to be of the order of unity,
and thus fix $\gamma=1.5$ and $\delta=-1$. This choice, which stems from the fact that
this ratio determines the constant density steady-state (see below),
will be particularly convenient for illustration purposes.
Furthermore, motivated by the fact that, in the context of optics,
parameters describing the higher-order effects take, typically, small values
\cite{KodHas87}, we fix $\mu=\nu=0.01$,
while third-order dispersion and SRS strengths, $\beta>0$, $\sigma_R>0$, are varied
in the intervals $[0,~1]$ and $[0,~0.3]$, respectively.

Obviously, the above choice is merely a low-dimensional projection of the full
6-dimensional parameter space. Nevertheless, since our scope here is to illustrate the
role of higher-order effects on the emergence of complex dynamics in Eq.~(\ref{eq1}),
we will show below that the variations of $\beta$ and $\sigma_R$ alone do
offer a clear physical picture in that regard.
To be more specific, the choice of those particular parameters stems from the following facts.
First, third-order dispersion is the sole linear higher-order effect, which is important
also in the linear regime (as it modifies the linear dispersion relation).
Second, the stimulated Raman scattering effect is the first higher-order
dissipative effect and, as such, is expected to play dominant role in the
long-time nonlinear dynamics of the system.

Naturally, the nontrivial task (as also highlighted above)
of investigating the full parameter space is interesting and relevant
in its own right,
yet it is beyond the scope of this work.


In our simulations, the limit set $\omega(u_0)$ will be visualized by projections
of the flow to suitable 2D or 3D spaces, defined by $\mathcal{P}_2=\left\{(X,Y)\in\mathbb{R}^2\right\}$,
and $\mathcal{P}_3=\left\{(X,Y,Z)\in\mathbb{R}^3\right\}$. Here,
$X(t)=|u(x_1,t)|^2$, $Y(t)=|u(x_2,t)|^2$, $Z(t)=|u(x_3,t)|^2$, for arbitrary spatial coordinates $x_1,\;x_2,x_3\in \Omega$. 

%
%

\begin{figure}[tbp]
	\centering
			\includegraphics[scale=0.24]{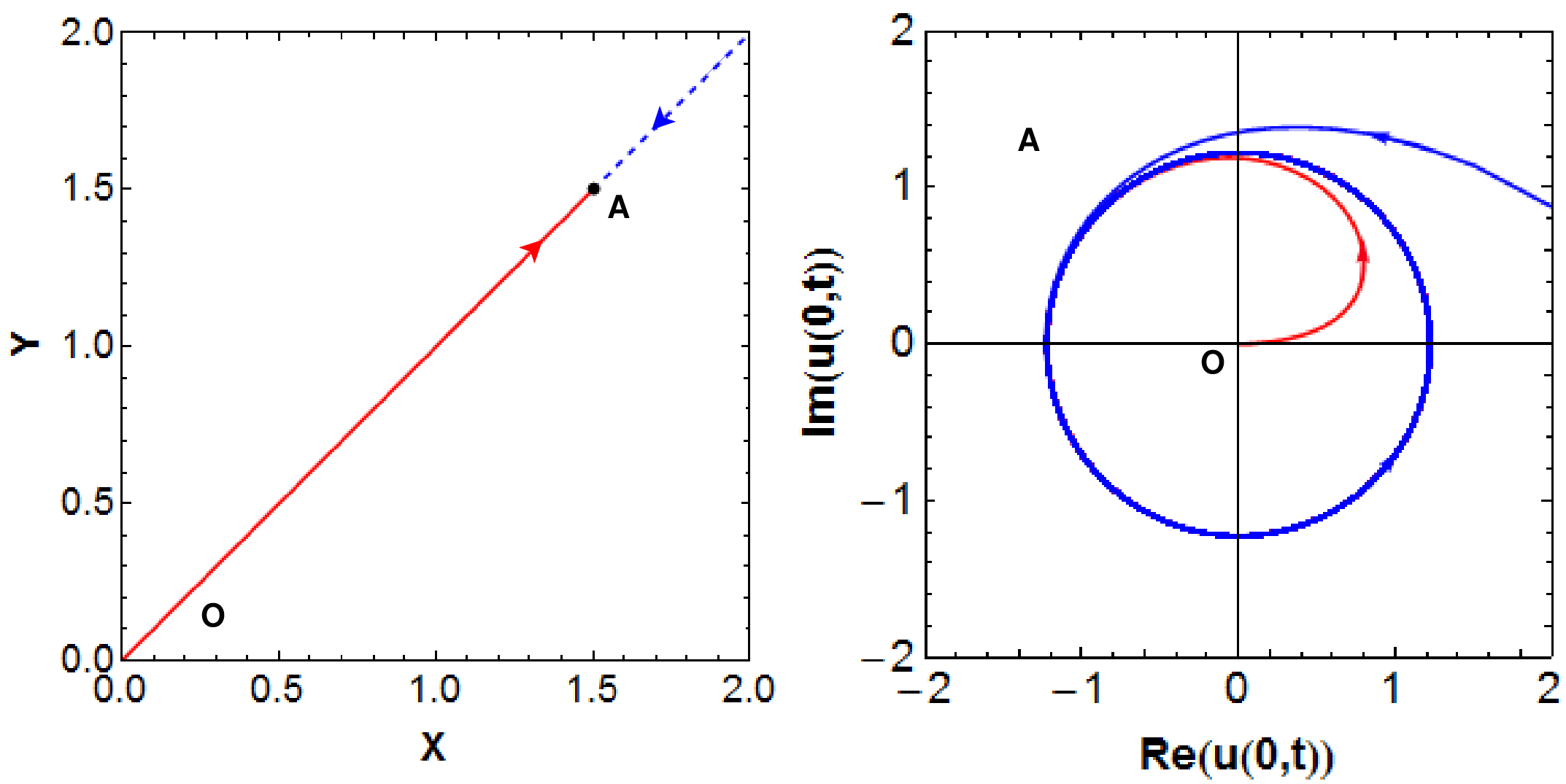}
		\caption{(Color Online) The scenario $\omega(u_0)=\{\phi_b\}$.
			Left panel: convergence to the fixed point $\mathbf{A}$.
			Right Panel: the fixed point $\mathbf{A}$ as a limit circle of radius
			$\sqrt{-\gamma/\delta}$.
			}
		\label{fig0}
\end{figure}

\subsection{Steady-state and orbital connections regime}

First, we use cw initial data,
$$u_0(x)=\epsilon\, \exp\left(-i\frac{K\pi x}{L}\right) \equiv \epsilon \phi_K$$
of amplitude $\epsilon>0$ and wave-number $K>0$,
which is an element of the 1D-linear subspace
$$\mathcal{V}_K=\left\{u\in L^2(\Omega)\;:\;u=\epsilon\phi_K(x),\;\;\epsilon>0\right\}$$
of $L^2(\Omega)$.
Here we should note that there exists a cw state which is an exact solution of Eq.~(\ref{eq1});
this solution is generically subject to modulational instability (MI) \cite{ostro}
(so-called Benjamin-Feir instability in the context of deep water waves \cite{BF}).
The exact cw solution, as well the relevant MI analysis are presented in Appendix B.
However, such analysis is not capable of providing any insights on the long-time dynamics
of the solutions. Indeed, although it can be used
as a means to understand the destabilization of the cw steady-state,
it does not offer any information regarding the long-time behavior
and the states the system passes through. As we show below,
the intricate dynamics that emerge, cannot be fully
understood in the framework of the MI picture.

Using the above cw initial data, and varying $\sigma_R>0$, 
we find that
$\omega(u_0)$ is an equilibrium state.
Specifically, there exists a critical wave number $K_{\mathrm{max}}$
%
such that: for $K <K_\mathrm{max}$, $\omega(u_0)=\phi_b$, i.e.,
a steady-state of constant density $|\phi_b|^2=-\frac{\gamma}{\delta}$;
for $K\geq K_\mathrm{max}$, 
$\omega(u_0)=\Phi_p$, i.e., a steady-state of spatially periodic density. We find that $K_{\mathrm{max}}$ decreases as $\sigma_R$ increases: if 
$\sigma_R=0,0.1,0.2,0.3$, and $\beta=0.02$, then $K_{\mathrm{max}}=16,13,10,5$, respectively.

The dynamical scenario $\omega(u_0)=\{\phi_b\}$ for $\beta=0.02$, $\sigma_R=0.3$ and  $K=4$ is illustrated in Fig.~\ref{fig0}. The projection of the cw equilibrium $\phi_b$ to the 2D space  $\mathcal{P}_2$ is the fixed point $\mathbf{A}=(|\phi_b|^2,|\phi_b|^2)=\left(-\frac{\gamma}{\delta},-\frac{\gamma}{\delta}\right)=(1.5,1.5)$. The right panel of Fig.~\ref{fig0} illustrates the convergence of the projected linear orbits to $\mathbf{A}$, associated to the choice of spatial coordinates $x_1=5, x_2=10$. The dashed blue (continuous red) line is the projection of the flow for the cw with $\epsilon=3$ ($\epsilon=0.01$); the arrows indicate the direction of the 2D-projection of the flow.
The cw steady state $\phi_b$ is an element of $\mathcal{V}_K$, and only differs in amplitude from the initial condition. Hence, $\mathcal{V}_K$ defines a stable linear subspace for $\mathbf{A}$. The right panel of Fig.~\ref{fig0} visualizes the steady state $\phi_b$ as a limit circle $\mathbf{A}$ of radius $\sqrt{-\frac{\gamma}{\delta}}=\sqrt{1.5}$, in the 2D space $(\mathrm{Re}(u(0,t)),\mathrm{Im}(u(0,t))$. The limit circle corresponds to the rotating linear oscillations of the real and imaginary parts of the solution $u$. Effectively in this case, the solution preserves
its plane-wave form but its amplitude, say $h(t)$, satisfies a Bernoulli equation.
This can be seen as follows: we substitute $u(x,t)=W(t) e^{\mathrm{i} K x}$
in Eq.~(\ref{eq1}) and obtain the following equation for the time-dependent amplitude $W(t)$ (as usual overhead dots denote time derivatives):
\begin{eqnarray}
\mathrm{i} \dot{W} - \frac{1}{2} K^2 W + |W|^2 W &=& \mathrm{i}\gamma W +\mathrm{i}\delta |W|^2 W
\nonumber \\
&+& \beta K^3 W -\mu K W.
\label{extra1}
\end{eqnarray}
Then, taking $W(t)= h(t) \exp[\mathrm{i} (K^2/2 - \beta K^3 + \mu K) t]$,
we obtain from Eq.~(\ref{extra1}) the Bernoulli equation:
%
%
%
$$\dot{h}=\gamma h+\delta h^3.$$
Thus, for $h(0)=\epsilon$,
$$\lim_{t\rightarrow\infty}h^2(t)=-\frac{\gamma}{\delta} \equiv |\phi_b|^2.$$
\begin{figure}[tbp]
\centering
\includegraphics[scale=0.26]{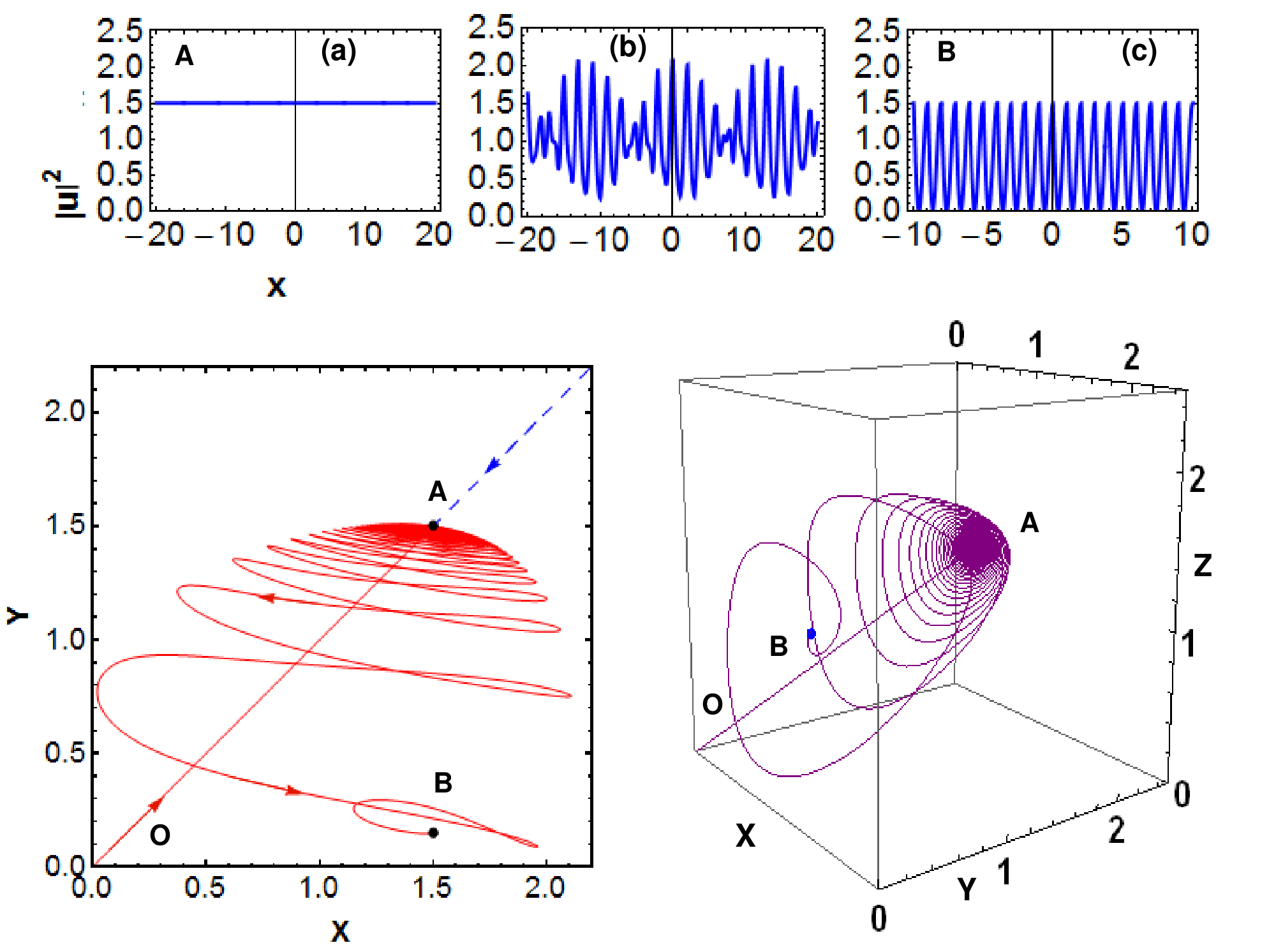}
		\caption{(Color Online) The scenario $\omega(u_0)=\{\Phi_p\}$.
			Upper panels: density snapshots at times (a) $t\approx500$, (b) $t\approx 683$, (c) $t\approx700$.
			Bottom panels:
			orbital connections $\mathbf{O}\rightarrow \mathbf{A}\rightarrow\mathbf{B}$ in 2D (left) and 3D (right) spaces.
			}
		\label{fig1}
\end{figure}

Next, consider the scenario $\omega(u_0)=\{\Phi_p\}$, for $\beta=0.02$ $\sigma_R=0.3$, and $K=5$, illustrated in Fig.~\ref{fig1}. The upper panel shows density snapshots for
a cw-initial condition with 
$\epsilon=0.01$. The solution has reached the cw-steady state $\phi_b$
exponentially fast, but at $t\approx 500$ the instability of the state $\phi_b$ emerges.
Although transient oscillations of increasing amplitude occur (cf. snapshot at $t=683$) due to the linear gain
$\gamma>0$, the nonlinear loss $\delta<0$ prevents collapse of the solution.
After $t\approx 685$, we observe convergence to the new steady state $\Phi_p$ (reached at $t\approx 700$), whose profile remains unchanged till the end of integration ($t=3000$).
The orbital connection, via
the transient dynamics, between steady states
$\phi_b$ and $\Phi_p$  is illustrated in the projections of the flow on the spaces $\mathcal{P}_2$ and $\mathcal{P}_3$ -- cf. bottom left and right panels of Fig.~\ref{fig1}, respectively, for $x_1=0$ and $x_2=4.5$. In 2D, $\mathbf{B}\approx (1.5,0.15)$
is the new fixed point, while in 3D,  $\mathbf{A}=(1.5,1.5,1.5)$ and
$\mathbf{B}\approx (1.5,0.15, 1.16)$.
The infinite-dimensional orbital connection:
$$\{0\}\;\;\mbox{(unstable)}\xrightarrow{\mathcal{O}_1}\{\phi_b\}\;\;\mbox{(unstable)}\xrightarrow{\mathcal{O}_2}\{\Phi_p\}=\omega(u_0),$$	
where $\mathcal{O}_1$ and $\mathcal{O}_2$ denote the orbits connecting the steady states, is projected to  the 2D and 3D-connections:
$$\mathbf{O}\;\;\mbox{(unstable)}\xrightarrow{\mathcal{O}'_1}\{\mathbf{A}\}\;\;\mbox{(unstable)}\xrightarrow{\mathcal{O}'_2}\{\mathbf{B}\}.$$
The projected orbits highlight the spiraling of the stable manifold of the limit point $\mathbf{B}$ around the unstable linear subspace of $\mathbf{O}=(0,0,0)$ connecting $\mathbf{O}$ and $\mathbf{A}$.  The connection was found to be stable with respect to variations of $\epsilon$ -- cf.
linear dashed blue (continuous red) converging orbit in the bottom left panel, corresponding to a cw-initial condition of amplitude $\epsilon=2$ ($\epsilon=0.01$).
\begin{figure}[tbp]
\centering
\includegraphics[scale=0.24]{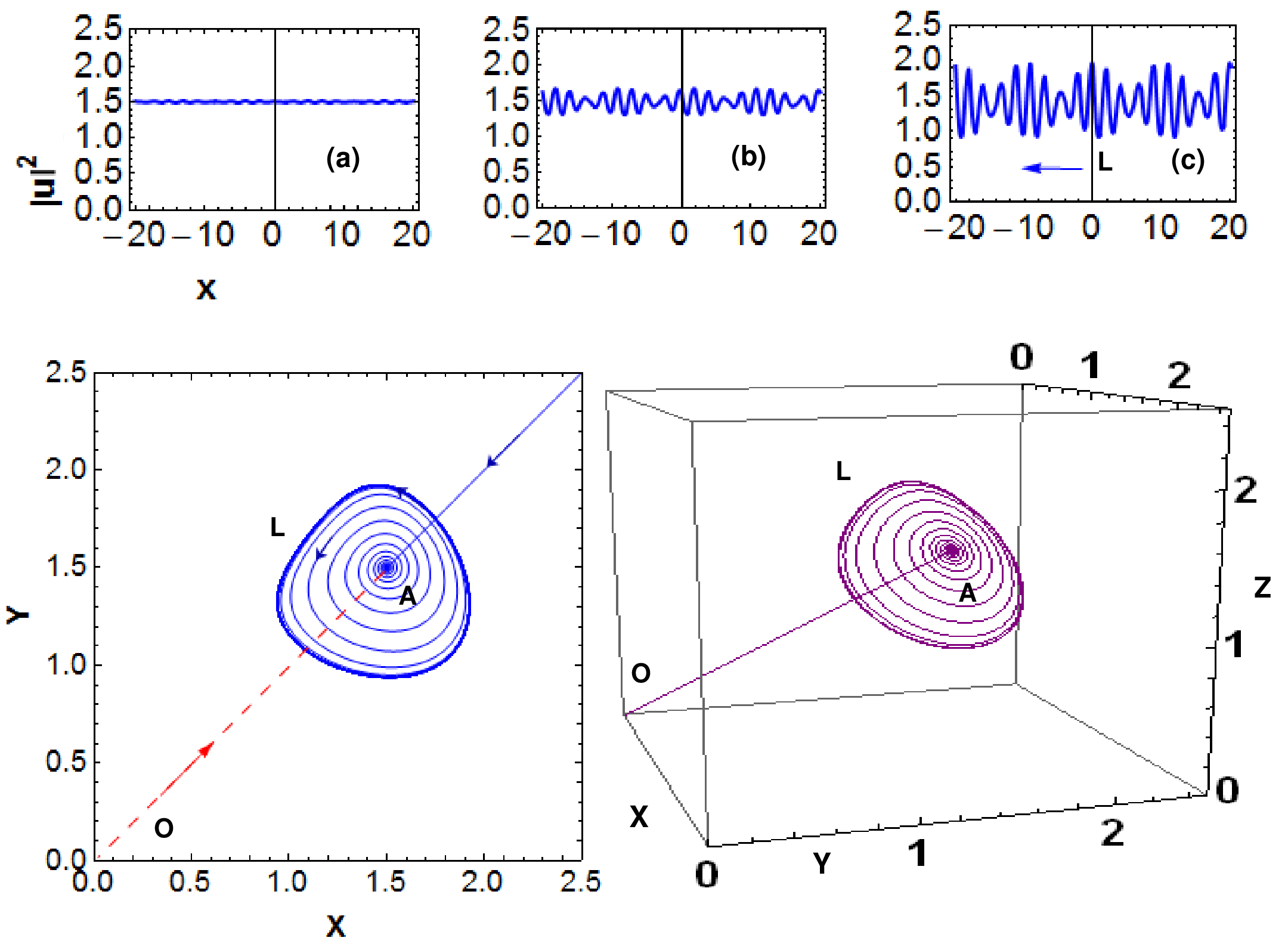}
		\caption{(Color Online) The dynamics scenario $\omega(u_0)=\mathbf{L}$, i.e., a space-time periodic traveling wave.
			Upper panels: density snapshots at times (a) $t\approx135$, (b) $t\approx 150$, (c) $t\approx 180$. 
			Bottom 
			panels: convergence $\mathbf{O}\rightarrow \mathbf{A}\rightarrow\mathbf{L}$, the limit cycle
			in 2D (left) and 3D (right) spaces.}
		\label{fig2}
\end{figure}


\subsection{Space-time periodic (limit-cycle) regime}

Increasing $\beta$, for $\sigma_R=0.01$, we observe the birth of yet another feature, namely traveling space-time oscillations. 
The upper panel of Fig.~\ref{fig2} shows density snapshots, for a cw initial condition
of $K=5$, $\epsilon=0.01$, and $\beta=0.55$. 
Now, instability
of the steady-state $\phi_b$, leads to the birth of a stable, traveling space-time periodic solution, whose profile is shown
for $t=180$ (arrow indicates propagation direction). The 
projections, 
for $x_1=0$, $x_2=5$ and $x_3=10$, on $\mathcal{P}_2$ (bottom left panel) and $\mathcal{P}_3$ (bottom right panel), visualize the periodic solution as a limit cycle $\mathbf{L}$, i.e., a periodic orbit.
The continuous blue (dashed red) linear orbit shown in the bottom left panel corresponds to the cw-initial condition of $K=5$ and $\epsilon=3$ ($\epsilon=0.01$), highlighting the stability (i.e., attracting nature)
of the limit cycle with respect to 
$\epsilon$.
\begin{figure}[tbp]
	\begin{center}
		\begin{tabular}{c}
			\hspace{-0.1cm}\includegraphics[scale=0.25]{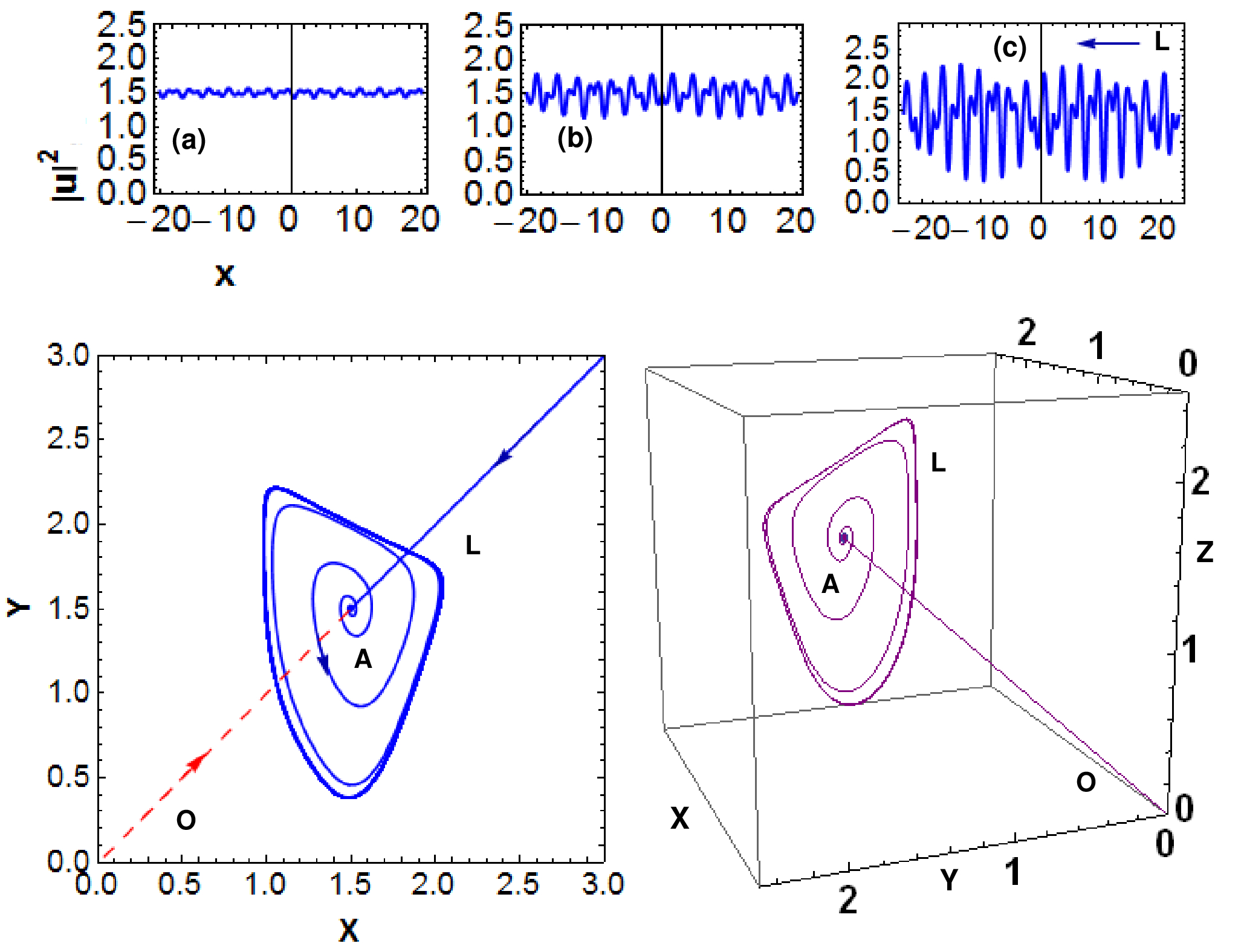}
		\end{tabular}
		\caption{(Color Online) 
The dynamics scenario $\omega(u_0)=\mathbf{L}$, in the presence of third-order dispersion only,
namely, for $\beta=0.02$ and $\mu=\nu=\sigma_R=0$. Upper panel: density snapshots at times (a) $t\approx 400$, (b) $t \approx 420$ and (c) $t \approx 450$, for a single-cw initial condition of $K=15$, $\epsilon=0.01$, 
			and $\beta=0.02$. Bottom left and right panels: convergence $\mathbf{O}\rightarrow \mathbf{A}\rightarrow\mathbf{L}$, i.e., the stable limit cycle, in the 2D and 3D spaces.}
		\label{fig2b}
	\end{center}
\end{figure}
 Specifically, 
for fixed $\sigma_R=0.01$ and $K>4$, there exists an interval $I_{\beta,K}=[\beta_{\mathrm{min}}(K), \beta_{\mathrm{max}}(K)]$, such that for some $\beta\in I_{\beta,K}$, the initial condition may converge to a space-time periodic solution; e.g., for $K=5$, 
$I_{\beta,5}\approx[0.5,0.57]$, while for $K=20$, $I_{\beta,20}\approx[0.7, 1.2]$.
On the other hand, when $\beta\notin I_{\beta, K}$, the initial condition  converges to a steady state. Evidently, the 
structure of the limit set $\omega(u_0)$ for Eq.~(\ref{eq1}), consisting either of distinct equilibria and orbits connecting them, or of a limit cycle, is reminiscent of
scenarios associated with PB dynamics.

It is important to remark that third-order dispersion plays a critical role in this scenario of
$\omega(u_0)=\mathbf{L}$, as it can be {\it solely} responsible for the emergence of a limit cycle.
Indeed, Fig.~\ref{fig2b} shows the dynamics for a cw-initial condition of $K=15$
and amplitudes as in Fig.~\ref{fig2}, but for $\beta=0.02$ and $\sigma_R=\mu=\nu=0$.
Furthermore, the third-order dispersion alone, can also give rise to even more complex
behavior (see below).

\begin{figure}[tbp]
\centering
			\includegraphics[scale=0.22]{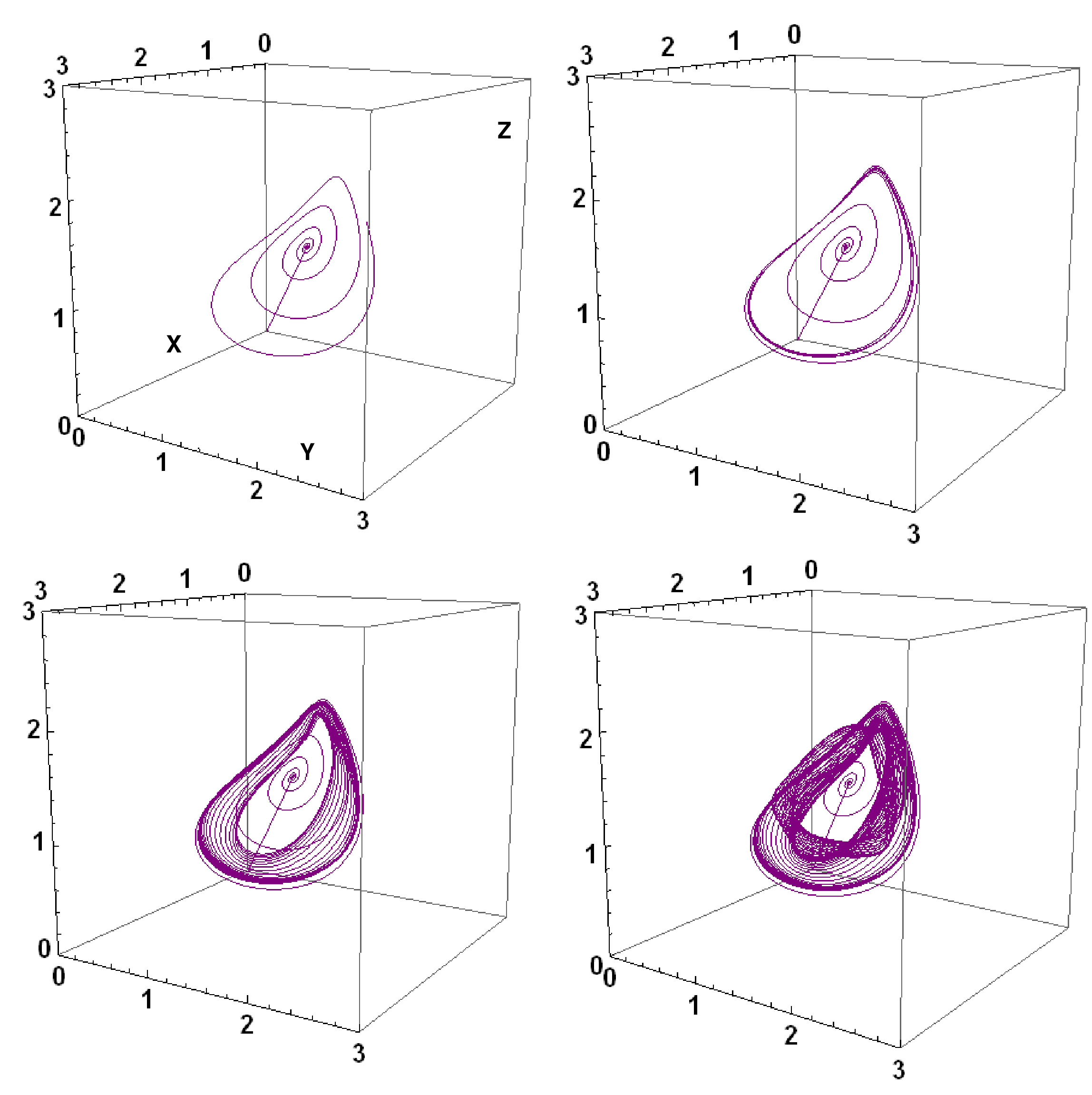}
		\caption{(Color Online) Birth of a chaotic attractor $\omega(u_0)=\mathbf{S}$.
		Transition from the instability of the cw-steady state $\mathbf{A}$, to quasiperiodic, and to chaotic behavior for $t\in [0,330]$.
}
		\label{fig3}
\end{figure}


\subsection{Quasi-periodic and chaotic regime}

The interval $I_{\beta,K}$ may be partitioned to sub-intervals where quasi-periodic, or even chaotic behavior emerges. Figure~\ref{fig3}
shows the 3D-projection of the flow on $\mathcal{P}_3$, for $x_1=5$, $x_2=10$, $x_3=15$, $t\in [0,350]$, 
$\beta=0.52$, $\sigma_R=\mu=\nu=0.01$, for a cw
of $\epsilon=0.01$ and $K=5$. We observe the birth of quasi-periodic orbits from the instability of the steady-state $\phi_b$, and the transition to chaotic behavior manifested by their trapping 
to a chaotic attractor $\mathbf{S}$.

\begin{figure}[tbp]
\centering
\includegraphics[scale=0.24]{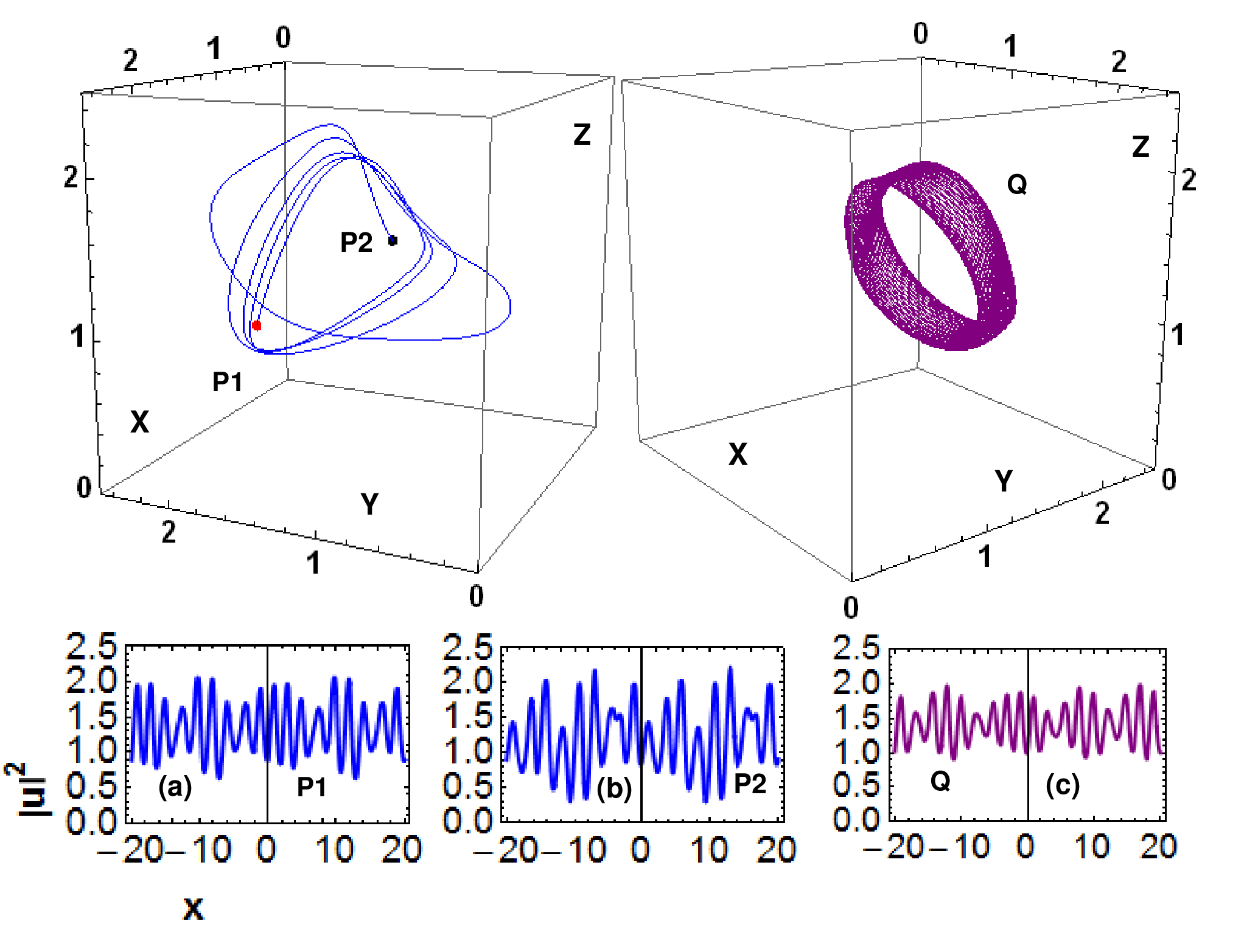} \vspace{0cm}
		\caption{(Color Online) Top left panel: a chaotic path in 
			$\mathbf{S}$ for $t\in [180,200]$.
			Top right panel: projection in 3D-space $\mathcal{P}_3$ of the invariant torus-like set $\mathbf{Q}$ for $t\in [1800,2000]$.
			Bottom panels: 
chaotic waveforms, corresponding to points $\mathbf{P1}$ at time $t\approx 150$ (left) and $\mathbf{P2}$ at time $t \approx 165$ (middle),
and a quasi-periodic solution in $\mathbf{Q}$  at time $t\approx 1900$ (right).
}
		\label{fig4}
\end{figure}

The upper left panel of Fig.~\ref{fig4} shows part of a chaotic orbit in $\mathbf{S}$, for $t\in [180,200]$, and  $\beta =0.5\approx \beta_{\mathrm{min}}(5)$.
The first two snapshots of the bottom panel show profiles of the solution corresponding to points $\mathbf{P1}$ and $\mathbf{P2}$, for $t=150$ and $t=165$.
The ``windings'' of the chaotic orbits are evident in the upper left panel of Fig.~\ref{fig4}, similarly also to the bottom right panel of Fig.~\ref{fig3}.
The chaotic behavior manifests itself in the time-fluctuating amplitude,
the changes in the waveform's spatial periodicity, and in the propagation direction of the chaotic traveling wave.

The interval $I_{\beta,K}=[\beta_{\mathrm{min}}(K), \beta_{\mathrm{max}}(K)]$ can be partitioned in the following sub-intervals: a chaotic  $I_{\beta,K,c}=[\beta_{\mathrm{min}}(K), \beta_{\mathrm{ch}}(K)]$, a quasi-periodic  $I_{\beta,K,q}=(\beta_{\mathrm{ch}}(K), \beta_{\mathrm{lc}}(K))$, and a limit-cycle one   $I_{\beta,K,lc}=[\beta_{\mathrm{lc}}(K), \beta_{\mathrm{max}}(K)]$. Let
$\beta_{\mathrm{min}}(K)$ be the critical value for the onset of the quasiperiodic behavior and the transition to the chaotic regime. Then, as $\beta\rightarrow \beta_{\mathrm{ch}}(K)$, the chaotic features are less evident and emerge at later times. Chaotic orbits still exist for $\beta=\beta_{\mathrm{ch}}(K)$. For $\beta>\beta_{\mathrm{ch}}(K)$, solutions remain quasi-periodic, and the orbit is trapped within an invariant torus-like set $\mathbf{Q}$. For $K=5$, we find that $\beta_{\mathrm{ch}}(5)\approx 0.53$.
The projection on $\mathcal{P}_3$ of
$\mathbf{Q}$ for $\beta=0.54>\beta_{\mathrm{ch}}(5)$, is shown in the upper right panel
of Fig.~\ref{fig4}. The orbit is plotted for $t\in [1800,2000]$, and the profile
of a quasi-periodic solution within $\mathbf{Q}$ at $t=1900$ is shown in the third snapshot of the bottom panel. The set $\mathbf{Q}$ persists as long as $\beta<\beta_{\mathrm{cl}}(K)$. When $\beta_{\mathrm{lc}}(K)\leq \beta\leq \beta_{\mathrm{max}}(K)$, the
set $\mathbf{Q}$ is replaced by a limit cycle. For 
$K=5$, we find the following sub-intervals of  $I_{\beta,5}\approx[0.5,0.57]$: the chaotic  $I_{\beta,5,c}\approx [0.5, 0.53]$, the quasi-periodic
$I_{\beta,5,q}\approx (0.53, 0.55)$, and the limit-cycle 
$I_{\beta,5,lc}\approx [0.55, 0.57]$. For 
$K=5$, the above sub-intervals were 
detected with accuracy $10^{-3}$: for $\beta=0.549$, the
set $\mathbf{Q}$ persists, while for $\beta=0.55$, the initial state 
is trapped on the limit cycle.

\subsection{Numerical bifurcation diagrams}

\begin{figure}[tbp]
\vspace{0.4cm}
	\centering
\hspace{-0.7cm}\includegraphics[scale=0.201]{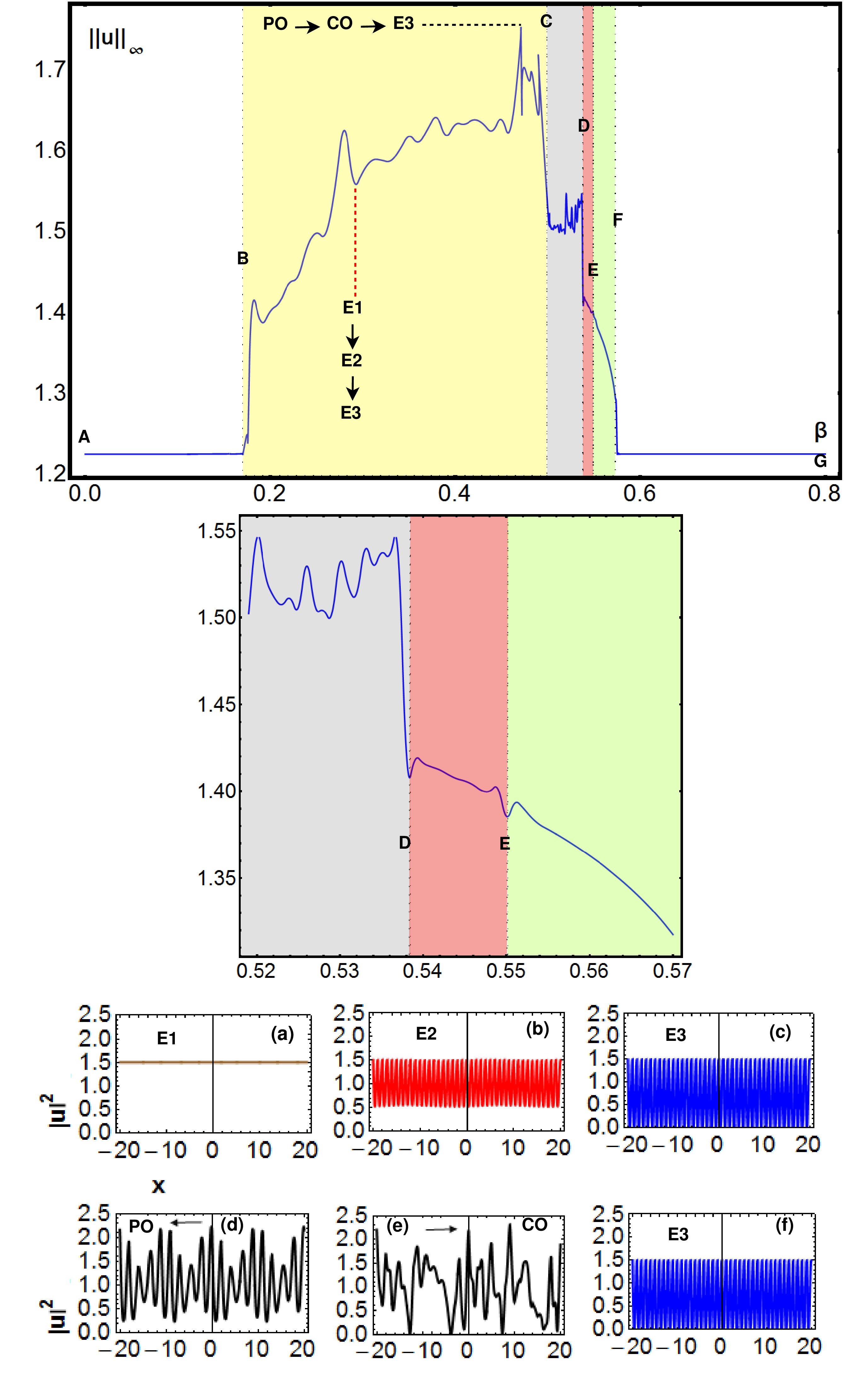}
	\caption{(Color Online) Top panel: $\beta- ||u||_{\infty}$ bifurcation diagram
	(Diagnostic~I), for fixed
$\sigma_R=\mu=\nu=0.01$, and the cw-initial condition of $\epsilon=0.01$ and $K=5$.
Second row panel: Magnification of the quasi-periodic region DE shown in the top panel.
Third row panels: Profiles of the distinct steady states involved in the orbital connection $\mathbf{E1}\rightarrow \mathbf{E2}\rightarrow\mathbf{E3}$ occurring at $\beta=0.3$ in the metastable region BC. The system is at rest in the steady-state $\mathbf{E1}$ for $5\lesssim t\lesssim 35$, in $\mathbf{E2}$ for $60\lesssim t\lesssim 68$, and in $\mathbf{E3}$ for $100\lesssim t\lesssim 3000$-the end of integration.
Fourth row panels: Transition from an unstable periodic orbit $\mathbf{PO}$ to chaotic oscillations $\mathbf{CH}$ which are eventually damped to the steady state $\mathbf{E3}$. The unstable periodic orbit  $\mathbf{PO}$ survives for $62\lesssim t\lesssim 105$, and the chaotic orbit for $120\lesssim t\lesssim 203$. The system is at rest in the steady state $\mathbf{E3}$ for $217\lesssim t\lesssim 3000$-the end of integration.  }
	\label{fig6}
\end{figure}

The richness of the dynamics can be summarized in
a bifurcation diagnostic (``diagnostic~I''), namely the
$\beta- ||u||_{\infty}$ bifurcation diagram,
shown in the upper panel of Fig.~\ref{fig6}. The bifurcation curve [continuous (blue) line]
illustrates the variations of the $||u||_{\infty}$-norm of the solutions, defined as
$$||u||_{\infty}={\rm max}_{(x,t)\in D}|u(x,t)|,~~D=[-L,L]\times [0, T_{\mathrm{max}}],$$
where $T_{\mathrm{max}}$ denotes
the end of the interval of numerical integration $[0,T_{\mathrm{max}}]$;
the third-order dispersion coefficient is $\beta\in [0,1]$, while the rest of parameters
are fixed to the values $\sigma_R=\mu=\nu=0.01$, and for the cw-initial condition we use
$\epsilon=0.01$ and $K=5$. The system was integrated until  $T_{\mathrm{max}}=3000$.
The branches AB and FG correspond to the intervals $\beta\in[0,0.18)$ and $\beta\in (0.57,1]$ respectively,
and are associated with the dynamical scenario $\omega(u_0)=\{\phi_b\}$, i.e., the convergence to the
steady-state of constant density, $|\phi_b|^2=-\frac{\gamma}{\delta}$.
The intersection
of the bifurcation curve with the auxiliary ``separatrix'' B, at $\beta\approx 0.18$,
designates the transition to the equilibrium metastability region BC [light grey (pale yellow) shaded area],
in the interval $\beta\in [0.18, 0.5)$. The fluctuations of the bifurcation curve are associated with
metastable dynamical scenarios between distinct states. One such scenario may refer to
the orbital connections between steady states mentioned above; another one, may correspond
to a transition from unstable periodic orbits to chaotic oscillations, and an eventual convergence
to a steady state. These scenarios are followed by drastically different transient dynamics
characterizing these connections.

As a first example, we note the metastable transition --
at $\beta=0.3$ [vertical dashed (red) line] -- between three
distinct steady-states $\mathbf{E1}\rightarrow \mathbf{E2}\rightarrow\mathbf{E3}$
(with $\mathbf{E1}$ marking the steady-state of constant density, $|\phi_b|^2=-\frac{\gamma}{\delta}$).
The third row panels of
Fig.~\ref{fig6} show density profiles of these steady states. A second example,
refers to the transition from an unstable periodic orbit $\mathbf{PO}$
(which emerges from the instability of the steady-state $\phi_b$), to chaotic oscillations
$\mathbf{CO}$ and the convergence to the final steady-state $\mathbf{E3}$; this transition
occurs for $\beta=0.47$ [horizontal dashed (black) line].
Density profiles during this transition are shown in the fourth row panels of Fig.~\ref{fig6}.
For the first example, the ultimate state
$\mathbf{E3}$ is reached at $t\approx 103$, and the solution remains unchanged
until the end of integration, while for the second example, the ultimate state
$\mathbf{E3}$ is reached at $t\approx 217$.

The intersection of the bifurcation curve with the
second auxiliary separatrix C, with an almost vertical slope, is associated with
the transition to the chaotic region CD (grey-shaded area), corresponding to the interval
$I_{\beta,5,c}\approx [0.5, 0.53]$. The sudden jump of the bifurcation curve (with an infinite slope)
at the intersection with the separatrix D designates the entrance into
the quasi-periodic regime DE [dark (pale red) shaded area],
associated with the interval $I_{\beta,5,q}\approx (0.53, 0.55)$. This region is magnified in the
second row panel of Fig.~\ref{fig6}.
On the other hand, the next steep jump at the intersection with the separatrix E
(also magnified in the second row panel of Fig.~\ref{fig6})
depicts the entrance to the space-time periodic regime EF [grey (pale green) shaded area],
associated with the limit-cycle interval $I_{\beta,5,lc}\approx [0.55, 0.57]$.
The limit-cycle branch bifurcates from the intersection with the separatrix F beyond which the
branch of the constant-density steady-state FG is traced.
\begin{figure}[tbp]
	\vspace{0.4cm}
	\centering
	\hspace{-0.52cm}\includegraphics[scale=0.201]{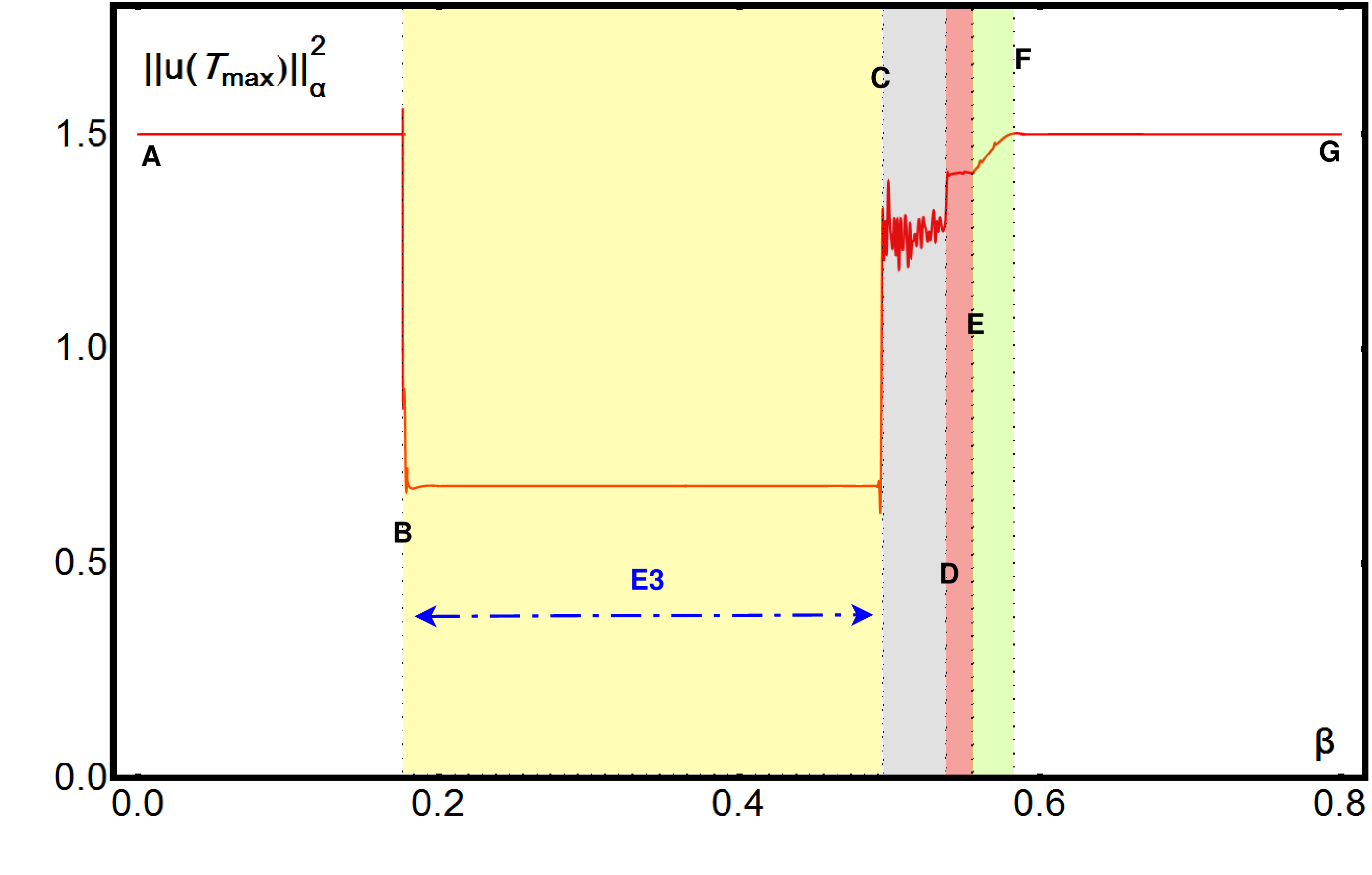}
	\caption{(Color Online) Top panel:
		$\beta- ||u(T_{\mathrm{max}})||^2_{\alpha}$ bifurcation diagram (Diagnostic~II), for fixed
		$\sigma_R=\mu=\nu=0.01$, and the cw-initial condition of $\epsilon=0.01$ and $K=5$.}
	\label{fig7.eps}
\end{figure}
Another bifurcation diagnostic (``diagnostic~II'') that we use herein, is the one associated with the variation
of the quantity
$$||u(T_{\mathrm{max}})||_{\alpha}^2=\frac{1}{2L}\int_{-L}^{L}|u(x,T_{\mathrm{max}})|^2dx$$
with respect to $\beta$.
For sufficiently large $T_{\mathrm{max}}$, $||u(T_{\mathrm{max}})||_{\alpha}^2$
could be thought of as
the superior limit of Eq.~(\ref{gees}).
The drawback in the above diagnostic is that the transient dynamics are hidden
(for sufficiently large $T_{max}$); more generally, the result strongly
hinges on the selection of $T_{max}$, but not necessarily strongly on the
evolution for earlier or mirroring that for later times.
Nevertheless, for sufficiently large $T_{max}$,
it can be particularly useful in detecting convergence to different steady-states,
e.g., $\omega(u_0)=\{\phi_b\}$ or $\omega(u_0)=\{\Phi_p\}$, via metastability.
Furthermore, it is also able to detect regimes of more complex behavior,
similarly to the $||u||_{\infty}$- diagnostic. Figure~\ref{fig7.eps} shows the
$\beta-||u(T_{\mathrm{max}})||_{\alpha}^2$ bifurcation curve [continuous (red) line],
for $T_{\mathrm{max}}=3000$; the rest of parameters are as in Fig.~\ref{fig6}.
The four shaded regions correspond to the same distinct dynamical regimes
that were detected in the $\beta- ||u||_{\infty}$ bifurcation diagram of Fig. \ref{fig6}.
The horizontal straight lines 
$$||u(T_{\mathrm{max}})||_{\alpha}^2=1.5=-\frac{\gamma}{\delta}$$
in the regions AB and FG show that, in these regimes of $\beta$,
solutions converge to the steady-state $\phi_b$. The intersection
of the bifurcation curve with the auxiliary ``separatrix'' B, at $\beta\approx 0.18$,
still designates the transition to the equilibrium metastability region BC.
However, the new horizontal straight line $||u(T_{\mathrm{max}})||_{\alpha}^2=0.68$
clearly shows that, after the transient metastability dynamics, the solution favors a particular steady-state of convergence, namely
$\mathbf{E3}$ for these parameters.

It is now useful to compare Diagnostics~I and II. First we note that
the comparison between the two
in the metastability regime BC, reveals that far-from-equilibria transient dynamics
are only identified by the fluctuations in the
$\beta-||u||_{\infty}$ curve (Diagnostic~I) -- and {\it not}
in the $\beta-||u(T_{\mathrm{max}})||_{\alpha}^2$ (Diagnostic~II).
%
These fluctuations can be understood by the fact that
$||u||_{\infty}$ may be reached at a certain instant, $t_0\in[0,T_{\mathrm{max}}]$ and also by
noting that,
in general,
$||u||_{\infty}\neq {\rm max}_{-L\le x\le L}|\Phi(x)|$
[i.e., the $||u||_{\infty}$-norm of a steady-state $\Phi(x)$].
Diagnostic~II, on the other hand, reveals that in the metastability regime BC, the dynamics
favors a distinct steady-state (as mentioned above) -- a fact that cannot be captured by
Diagnostic~I.
\begin{figure}[tbp]
	\vspace{0.4cm}
	\centering
	\hspace{-0.4cm}\includegraphics[scale=0.201]{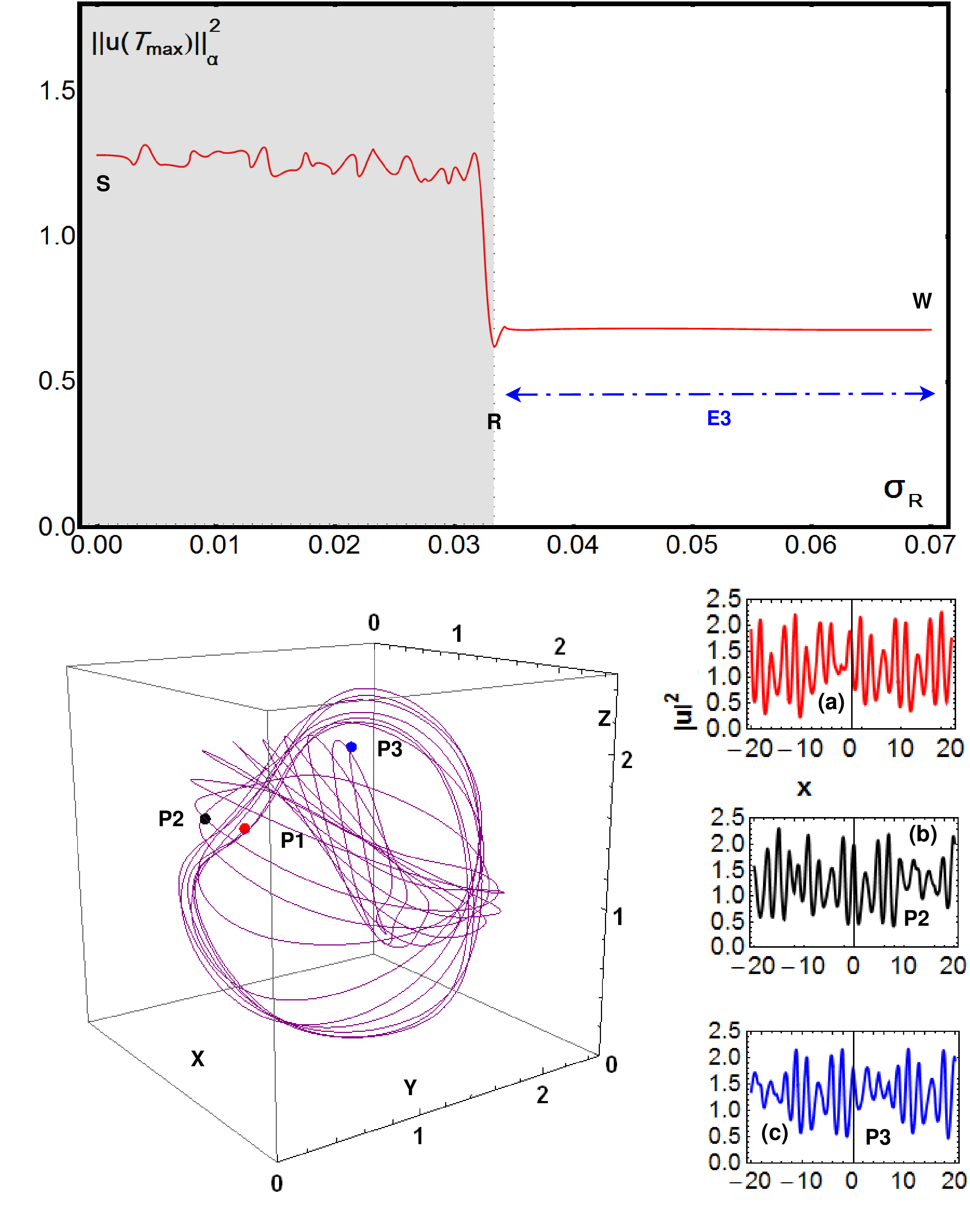}
	\caption{(Color Online) Top panel:
		$\sigma_R- ||u(T_{\mathrm{max}})||^2_{\alpha}$ bifurcation diagram (Diagnostic~II), for fixed
		$\beta=0.52$, $\mu=\nu=0.01$, and the cw-initial condition of $\epsilon=0.01$ and $K=5$. Bottom left panel: A chaotic path for $t\in[600, 650]$, when $\beta=0.53$, $\sigma_R=\mu=\nu=0$ and the initial condition is as in the top panel. Bottom right panel: chaotic waveforms corresponding to points $\mathbf{P1}$ at $t \approx 600$ (top), $\mathbf{P2}$ at $t\approx 625$ (middle), and $\mathbf{P3}$ at $t\approx 650$. }
	\label{fig8}
\end{figure}
As far as the other regimes are concerned, Diagnostic~II
can also capture the transition to the chaotic regime CD, indicated by
the intersection of the bifurcation curve with the auxiliary separatrix C, as well as
by its large rapid fluctuations
within region CD. The sudden jump of the bifurcation curve
at the intersection with the separatrix D designates the entrance into
the quasi-periodic regime, portrayed by the small, almost horizontal branch
of quasi-periodic solutions within region DE.
Note that the transition to the quasi-periodic regime is much more apparent
in the Diagnostic~II than in Diagnostic~I.
The intersection of the bifurcation curve with the separatrix E (at a point where the curve has a
local minimum in the region DF), is again associated
with the entrance to the space-time periodic regime EF (corresponding to the
branch of space-time periodic solutions). This branch
bifurcates from the straight line FG (pertinent to constant density steady-states)
at its intersection with the separatrix F.

It is important to make, at this point, yet some additional remarks.
First, the interval $I_{\beta,K}$, corresponding to the region CF in the bifurcation diagrams,
was found to be unstable under variations of $\sigma_R>0$. Corresponding (in)nstability
regimes are illustrated in the top panel of Fig.~\ref{fig8}, where a Diagnostic~II-type diagram
is shown, namely the bifurcation curve $\sigma_R-||u(T_{\mathrm{max}})||_{\alpha}^2$
[continuous (red) line]. This diagram is plotted for fixed $T_{\mathrm{max}}=3000$ and $\beta=0.52$
(recall that, in the previous case, for fixed $\sigma_R=0.01$, it was found that
$\beta=0.52\in I_{\beta,5,c}\approx [0.5, 0.53]$, i.e., in the chaotic regime);
the rest of parameters are as in Fig.~\ref{fig6}. It is observed that
for relatively small values of the SRS coefficient, namely for $\sigma_R<0.03$
(cf. grey-shaded area, labeled by SR), chaotic behavior persists. On the other hand,
above this threshold, i.e., for $\sigma_R> 0.03$, chaotic structures are destroyed,
and the system enters into the metastability regime (labeled by RW in the diagram).
The ultimate steady-state is $\mathbf{E3}$ for these parameters.
Note that the instability of quasi-periodic and space-time periodic regimes under the influence of
small increments of $\sigma_R$, occurs in a very similar manner, and can be plotted in similar
bifurcation diagrams (results not shown here).

Second, the interval $I_{\beta,K}$ persists even in the absence
of the rest of the higher-order effects, i.e., for $\sigma_R=\mu=\nu=0$.
This highlights the fact that the third-order dispersion 
plays a dominant role in the emergence of complex dynamics. An example of the chaotic behavior,
for $\beta=0.53$ and $\mu=\nu=\sigma_R=0$, is shown in the bottom panels of Fig.~\ref{fig8}.
In particular, the bottom left panel shows a part of a chaotic orbit for $t\in [600,650]$,
of the 3D-projection of the flow on $\mathcal{P}_3$, for $x_1=5$, $x_2=10$, and $x_3=15$.
Furthermore, the three snapshots in the bottom right panel, show profiles of the solution
corresponding to points $\mathbf{P1}$, $\mathbf{P2}$, and $\mathbf{P3}$ of the chaotic path
shown on the left, for $t=600$, $t=625$, and $t=650$, respectively.


\section{Conclusions}

In conclusion,
we have studied a physically important and broadly relevant higher-order
Ginzburg-Landau equation, with zero diffusion. The considered model, 
is motivated by a higher-order nonlinear Schr{\"o}dinger equation, which finds 
applications in a variety of contexts, ranging from nonlinear fiber optics and 
deep water waves; the model also incorporates linear loss and nonlinear gain, 
while it is supplemented with periodic boundary conditions, which are relevant 
to optical cavities settings, such as ones employed, e.g., in ring lasers.

Our analysis revealed that the infinite-dimensional dynamics of
this model
can be reduced 
to a sequence of low-dimensional dynamical scenarios
(fixed points, periodic and quasi-periodic, as
well as chaotic orbits) that can be suitably revealed in reduced
(two- and three-dimensional) phase space representations.
%
Such a dynamical picture is shared by various non-integrable perturbations of Hamiltonian partial
differential equations (such as the NLS, Sine-Gordon, and others), as these perturbations
may break the homoclinic structure of their integrable counterparts. However, the path
to all the above dynamical scenarios can be traced in drastically different ways and to essentially distinct roots,
even if the systems have similar origins for their dissipative nature manifested by the existence of an attractor, e.g., due to the presence of gain/loss, as in the case of CGL models.

In particular, in our higher-order CGL model, keeping gain/loss -- as well
as other coefficients of the higher-order effects -- fixed, we have shown that the competition between
third-order dispersion and the SRS effect (in the presence of
nonlinearity, dispersion, and gain/loss) can trace a path from Poincar\'{e}-Bendixson--type behavior
to quasi-periodic or chaotic dynamics. These dynamical transitions
are also reminiscent of ones 
observed in fiber ring lasers, or in the path towards optical turbulence phenomena \cite{VanWig,Ikeda}. A conspicuous finding was that third-order dispersion chiefly appears to be playing a critical role in controlling
the transition from periodic
to quasi-periodic, and  eventually to chaotic behavior, even in the absence of the rest of the
higher-order effects.

Our results highlight that higher-order effects may have a primary role for the birth of spatiotemporal  transitions in mixed gain/loss systems, suggesting further investigations.
First of all, in the framework of the model we considered herein, it would be particularly
interesting to investigate more broadly the full six-parameter space, rather than its low-dimensional
projection considered herein. Furthermore, another interesting direction would be
the identification of a low-dimensional attractor, its dimension
and dependence on the spatial length \cite{eli}, as well as the construction
of the appropriate finite-dimensional reduced systems
able to capture the effective low dimensional dynamics \cite{Cv2}.
Lastly, it would
also be interesting to investigate the role of higher-order effects
in other autonomous systems with gain and loss.
%
%
\appendix
\section{Existence of a limit set (attractor)}
\label{apLS}
In this Appendix, we define  an extended dynamical system associated to the initial-boundary value problem (\ref{eq1})-(\ref{eq3}). In particular, we briefly sketch the proof for the existence of a limit set-attractor, capturing all bounded orbits of this dynamical system, which initiate from sufficiently smooth initial data (\ref{eq3}).

The starting point of our proof is the power balance equation 
\cite{remark}:
\begin{eqnarray}
\label{OL2ge}
\frac{d}{dt}\int_{-L}^{L}|u|^2dx=2\gamma\int_{-L}^{L}|u|^2dx
+2\delta\int_{-L}^{L}|u|^4dx,
\end{eqnarray}
satisfied by any local solution $u\in C([0,T], H^k_{per}(\Omega))$, which  initiates from sufficiently smooth initial data $u_0\in  H^k_{per}(\Omega)$, for fixed $k\geq 3$. Here, $H^k_{per}(\Omega)$ denotes the Sobolev spaces of periodic functions $H^k_{per}$ \cite{RTem88}, in the  fundamental interval $\Omega=[-L,L]$: 
\begin{eqnarray*}
H^k_{per}(\Omega)=\{u:\Omega\rightarrow \mathbb{C},\,\,u\,\mbox{and}\,\, \partial^j_xu\in L^2(\Omega),\,
j=1,...,k;\nonumber\\ 
u\;\mbox{and}\;\partial^j_xu\;
\mbox{for $j=1,...,k-1$, are $2L$-periodic}
\}.
\end{eqnarray*}

Analysis of (\ref{OL2ge}), results in the asymptotic estimate:
\begin{eqnarray}
\label{gees}
\limsup_{t\rightarrow\infty}
\frac{1}{2L}\int_{-L}^{L}|u(x,t)|^2dx\leq-\frac{\gamma}{\delta},
\end{eqnarray}
hence local in time solutions $u\in C([0,T], H^k_{per}(\Omega))$ are uniformly bounded in $L^2(\Omega)$. This allows for the definition of the extended dynamical system
$$\varphi(t, u_0): H^k_{per}(\Omega))\rightarrow L^2(\Omega),~~\varphi(t, u_0)=u,$$
whose orbits are bounded $\forall t\geq 0$. Moreover, from the above asymptotic estimate, we 
derive, that if $L^2(\Omega)$ is endowed with the equivalent averaged norm
$$||u||^2_{\alpha}=\frac{1}{2L}\int_{-L}^{L}|u|^2dx$$
then its ball
$$\mathcal{B}_{\alpha}(0,\rho)=\left\{u\in L^2(\Omega)\;:\;||u||^2_{\alpha}\leq \rho^2,\;\;\rho^2>-\frac{\gamma}{\delta}\right\}$$
attracts all bounded sets $\mathcal{B}\in H^k_{per}(\Omega)$. That is,
there exists $T^*>0$, such that
$\varphi(t, \mathcal{B})\subset \mathcal{B}_{\alpha}$, for all $t\geq T^*$.  Thus, we may define for any bounded set $\mathcal{B}\in H^k_{per}(\Omega))$, $k\geq 3$,  its $\omega$-limit set in $L^2(\Omega)$,
$$\omega(\mathcal{B})=\bigcap_{s\geq 0}\overline{\bigcup_{t\geq s}\varphi(t ,\mathcal{B}}).$$
The closures are taken with respect to the weak topology of $L^2(\Omega)$. Then, the standard (embedding) properties of Sobolev spaces imply that the attractor $\omega(\mathcal{B})$ is at least weakly compact in $L^2(\Omega)$, or relatively compact in the dual space $H^{-1}_{per}(\Omega)$.  For any initial condition (\ref{eq3}), $u_0\in\mathcal{B}$, we denote its limit set by $\omega(u_0)\subset \omega(\mathcal{B})$.
\appendix
\setcounter{section}{1}
\label{apMI}
\section{Modulational instability}

In this Appendix we provide the modulational instability analysis of the cw state:
%
%
\begin{equation}\label{cw}
u=u(t)=A e^{{\rm i}\theta(x,t)},\quad \theta(x,t)= k_0 x-\omega_0 t,
\end{equation}
(where $A$ is a real constant), which is an exact analytical solution of Eq.~(\ref{eq1})
(for a MI analysis for the cw solution of Eq.~(\ref{eq0}) cf. Ref.~\cite{potasek}).
This solution exists when the following dispersion relation holds:
$${\omega _0} = {\beta k_0^3 - k_0^2/2 - \mu {A^2}{k_0} + {\rm i}\left( {\gamma  + \delta {A^2}} \right) - {A^2}},$$
while $A^2=-\gamma/\delta$, to suppress any exponential growth.
This amplitude value is consistent with the equilibria (steady states) of the system.

Now consider a small perturbation to this cw solution
\[
u(x,t)=[A + u_1(x,t)]e^{{\rm i}\theta(x,t)},
\]
inserted into Eq.~\eqref{eq1}. Linearizing the system with respect to $u_1$ we obtain
\begin{align*}
{\rm i}({u_{1t}}-k_0 u_{1x}) &- \frac{1}{2}{u_{1xx}}+ A^2(u_1 + u_1^*)=  {\rm i}\delta A^2 (u_1+u_1^*)\\
&+ {\rm i}\beta ( 3k_0^2{u_{1x}} - 3{\rm i}{k_0}{u_{1xx}} - {u_{1xxx}}) \\
&- {\rm i}\mu A^2({\rm i}{k_0}{u_1} + {\rm i}{k_0}u_1^* + 2{u_{1x}} + u_{1x}^*) \\
&- {\rm i}(\nu-{\rm i}\sigma_R) A^2({u_{1x}}+u_{1x}^*),
\end{align*}
where star denotes complex conjugate.
Solutions of the above equations are sought in the form:
\[
u_1(x,t)=c_1 e^{i (k x-\omega  t)} + c_2 e^{-i (k x-\omega  t)},
\]
where $c_{1,2}$ are real constants, while $k$ and $\omega$ are the wavenumber and frequency
of the perturbations. This way, we obtain the dispersion relation:
\begin{equation}\label{disp}
  \delta^2\omega^2+p_1(k)\omega+p_2(k)=0
\end{equation}
where
\begin{align*}
p_1(k) &=  - 2\beta {k^3} + 2[ - 3\beta k_0^2 + {k_0} + {A^2}(2\mu  + \nu  - {\rm i}\sigma_R )]k, \\
p_2(k) &=
  {\beta ^2}{k^6} \\&+ [ - 3{\beta ^2}k_0^2 + \beta {k_0}
  - 2\beta {A^2}(2\mu  + \nu  - {\rm i}\sigma_R ) - 1/4]{k^4} \\&+
  [9{\beta ^2}k_0^4 - 6\beta k_0^3 + k_0^2( {1 - 6\beta {A^2}(\mu  + \nu  - {\rm i}\sigma_R )} ) \\&+ {k_0}{A^2}(\beta (6 - 6{\rm i}\delta ) + 3\mu  + 2\nu  - 2{\rm i}\sigma_R ) \\&+ {A^2} ( {{\rm i}\delta  + \mu {A^2}(3\mu  + 2\nu  - 2{\rm i}\sigma_R ) - 1}]{k^2},
\end{align*}
and it should be recalled that $A^2=-\gamma/\delta$. It is clear that
the system will always be modulationally unstable,
since the solutions of Eq.~\eqref{disp} are in general complex.
%


\begin{thebibliography}{99}
\bibitem{remark} 
To obtain the equation, multiply Eq. \eqref{eq1}
by $u^*$ and its conjugate by $u$, add the equations and integrate in $[-L,~L]$ using the relevant boundary conditions.
\end{thebibliography}
\end{document}